\documentclass{iopart}

\usepackage{amssymb}
\usepackage{graphicx}
\usepackage{enumerate}

\newcommand {\be} {\begin{equation}}
\newcommand {\ee} {\end{equation}}
\newcommand {\bea} {\begin{eqnarray}}
\newcommand {\eea} {\end{eqnarray}}
\newcommand{\tq}{\tilde{q}}
\newcommand{\tom}{\tilde{\omega}}
\newcommand{\Lie}{\mathcal{L}}
\newcommand{\tep}{\tilde{\epsilon}}
\newcommand{\tOm}{\tilde{\Omega}}

\newcommand{\tge}{{\Gamma}_{ex}}

\begin{document}

\title{Some spacetimes containing non-rotating extremal isolated horizons}

\author{Ivan Booth$^1$ and David Wenjie Tian$^2$}
\address{$^1$ Department of Mathematics and Statistics \\ Memorial University of Newfoundland \\ 
St. John's, NL,  A1C 5S7,  Canada} 
\address{$^2$ Department of Physics and Physical Oceanography \\ Memorial University of Newfoundland \\
 St. John's, NL,  A1B 3X7,  Canada}
 \ead{ibooth@mun.ca,  wtian@mun.ca}



\pacs{04.20.-q, 04.20.Ex, 04.20.Jb, 04.70.Bw}


\begin{abstract}
Well-known results demonstrate the uniqueness of extremal isolated horizons (equivalently near-horizon spacetimes) in $(3+1)$-dimensions. 
This paper briefly reviews some of these results and then explicitly constructs families of non-asymptotically flat, non-spherically symmetric spacetimes that nevertheless 
contain spherically symmetric extremal horizons that are isomorphic to those found
in Reissner-Nordstr\"om spacetimes. 
\end{abstract}

\section{Introduction}

The black hole uniqueness theorems are classic results in general relativity. It has been 45 years since Israel demonstrated
that Schwarzschild and Reissner-Nordstr\"om are respectively the unique (asymptotically flat) static black hole solutions to the 
$(3 \! + \! 1)$-dimensional 
Einstein and Einstein-Maxwell equations\cite{Israel:1967wq,Israel:1967za}, 40 since Kerr was shown to be the unique (asymptotically flat) 
stationary vacuum black hole solution \cite{carter, robinson} and 30 since the analogous (non-extremal) Kerr-Newman electrovac result was 
proved\cite{mazur}. More recently these results have been extended to extremal black holes \cite{Amsel:2009et} as well as considered 
in higher dimensions (for a review see \cite{Hollands:2012cc}). 

Those results are for full spacetimes. Over the last decade, a new set of uniqueness and classification results have been proved which focus on
the horizon alone. In particular it has been realized that extremal horizons must satisfy a particularly  tight set of constraint equations. 
These have been employed in both the isolated horizon \cite{Lewandowski:2002ua,Lewandowski:2006mx} and near-horizon 
(for example \cite{hariOld, hariClass,Hollands:2009ng,hariNewer}) formalisms and were anticipated in \cite{hajicek}. 

Though certainly non-trivial, working with these constraints is significantly easier than dealing with the full spacetime problem: instead of the 
Einstein equations on an $(n\!+\!1)$-dimensional Lorentz signature spacetime, the simpler constraint equations are formulated on 
$(n\!-\!1)$-dimensional Riemannian cross-sections of the horizon. As such this approach has enabled significant progress in constraining 
and classifying the range of extremal black holes not only in $(3\!+\!1)$ but also in higher dimensional spacetimes.

The two classes of uniqueness theorems complement each other but are not equivalent. 
The horizon-centred methods generate the set of solutions to the horizon constraints but do not demonstrate whether or not those horizons are
actually realized as parts of full solutions to the Einstein equations; existence must be demonstrated separately.
The global methods work directly with full spacetimes, but in doing that include extra constraints on the asymptotic
structure and possible matter fields. There is then a space to explore between the theorems: are there spacetimes that
contain Kerr-Newman-type extremal isolated horizons but violate the global constraints? One would expect the answer to be
yes but this then leads to a second question: what do those spacetimes look like and do they have to share the symmetry properties
of the horizon? For example, can a spherically symmetric extremal horizon live in a non-spherically symmetric spacetime?

%

In this paper we explicitly demonstrate that there are non-asymptotically flat, 
non-spherically symmetric spacetimes which contain (spherically symmetric) 
Reissner-Nordstr\"om-type extremal horizons\footnote{For a nice discussion of deformations of extremal Kerr horizons, see \cite{dainKerr}. }. To do this we 
examine Weyl\cite{gautreau,Geroch:1982bv,fk} and conformastatic\cite{Gonzalez:2008rn, LoraClavijo:2010ih, conformastatic} 
distorted Reissner-Nordstr\"{o}m spacetimes. 
The resulting spacetimes are static but neither spherically symmetric nor asymptotically flat. However they still contain the 
original spherically symmetric horizon. This contrasts strongly with the non-extremal case where Weyl distortions of the full spacetime can
dramatically deform the horizon geometry \cite{fk,frolov1,frolov2,Pilkington:2011aj}.



In outline we proceed as follows. Section \ref{background} reviews the geometry of stationary horizons, the constraint equations that they must satisfy and 
the uniqueness theorems for extremal horizons.  
Section \ref{Examples} reviews Weyl and conformastatic distortions, 
applies them to extremal black hole spacetimes and so constructs families of (non-asymptotically flat) spacetimes sharing a given 
extremal horizon. Section \ref{discuss} concludes with some reflections on these results.
\ref{AppA} provides details of how the Weyl form of Reissner-Nordstr\"om 
is transformed into the standard form.

\section{Background and Set-up}
\label{background}

We consider $(3\!+\!1)$-dimensional electrovac spacetimes without a cosmological constant.
 Then our spacetimes will solve
\bea
 \mathcal{R}_{ab} 
= 8 \pi T_{ab} \, ,  \label{Einstein}
\eea
where the (trace-free) stress-energy tensor takes the form:
\bea
T_{ab} = \frac{1}{4 \pi} \left(F_{ac} F_b^{\phantom{b}c} - \frac{1}{4} g_{ab} F_{cd} F^{cd}  \right) \, . \label{T}
\eea
As usual the electromagnetic field tensor $F_{ab}$ satisfies the source-free Maxwell equations 
\bea
\nabla_a F^{ab} = 0  \; \mbox{and} \; \nabla_{[a} F_{b c]} = 0 \, , \label{Maxwell}
\eea
where the square brackets indicate the usual anti-symmetrization. 

\subsection{Horizon geometry}
For spacetimes with asymptotic structures that are neither flat nor AdS, event horizons are not well defined. The spacetimes of 
Section \ref{Examples} have such non-standard structures and as such their horizons are not event horizons. Instead we 
consider null surfaces with time-invariant geometries. These may be viewed as either isolated horizons\cite{afk,abl,abl2} or Killing horizons 
(for example, \cite{frolov}) according to taste: our horizons will all live in static spacetimes and so the distinction between time-invariant horizons
(isolated horizons) and time-invariant horizons which also have time-invariant spacetime neighbourhoods (Killing horizons) is not significant. 

%
%

\begin{figure}
\begin{center}
\includegraphics{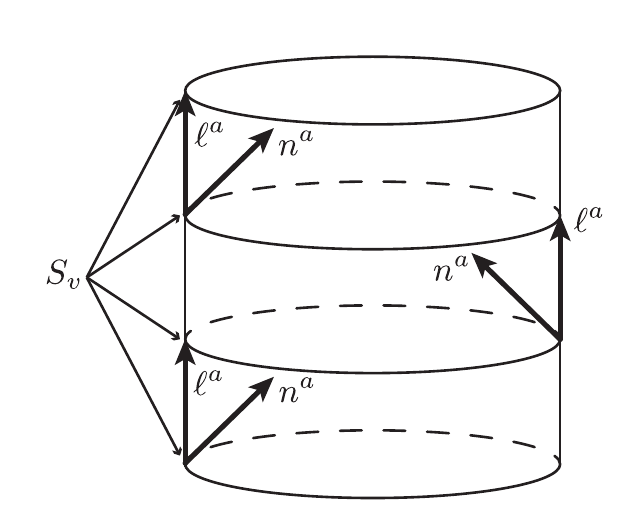}
\end{center}
\caption{The horizon $\Delta$ along with its foliation and the null normals to $S_v$.}
\label{Fig}
\end{figure}

It is sufficient to focus on three-dimensional null surfaces $\Delta$ which can be foliated into surfaces $S_v$, each of which is 
diffeomorphic to $S^2$. The normal one-form to $\Delta$ is $\ell_a$ and it is oriented and scaled so that $\ell^a$, which is tangent to the horizon, 
is future-pointing with $\Lie_\ell v = 1$. The other null normal to the $S_v$ is $n_a$. It is oriented so that $n^a$ is future-pointing and scaled so that
$\ell \cdot n = -1$. Thus $n = dv$ in $T^\star \! \Delta$.  Figure \ref{Fig} presents a schematic of $\Delta$ and these vector fields. 

We describe the geometry of $\Delta$ starting with the individual $S_v$. 
The intrinsic geometry of those surfaces is defined by the induced metric 
\bea
\tq_{AB} = e_A^a e_B^b g_{ab} 
\eea
where $e_A^a$ is the pullback operator from $T^\star \! M$ to $T^\star \! S_v$, while the extrinsic geometry is determined by the extrinsic curvatures relative to 
$\ell_a$ and $n_a$
\bea
\tilde{k}^{(\ell)}_{AB} = e_A^a e_B^b \nabla_a \ell_b  \; \; \mbox{and} \; \; \tilde{k}^{(n)}_{AB} = e_A^a e_B^b \nabla_a n_b \, ,
\eea
as well as the connection on the normal bundle
\bea
\tom_A = - e_A^a n_b \nabla_a \ell^b \, . 
\eea
These fully define the geometry of the $S_v$ and, in fact, if we add in the inaffinity $\ell^a$: 
\bea
\kappa_{(\ell)}= - n_a \ell^b \nabla_b \ell^a \,, 
\eea
they are sufficient to determine the full geometry of $\Delta$. The notation used here follows that of \cite{bfbig} quite closely. In particular tildes are used to 
flag quantities defining the geometry and electromagnetic properties of the $S_v$.

Physically, $\tilde{\omega}_A$ is often referred to as the angular momentum one-form. It (or sometimes a closely related quantity) is integrated against a rotational symmetry 
generating vector-field in most popular definitions of angular momentum (an extended discussion of this point can be found in  \cite{bfbig}). The curvature of this connection 
is $\tilde{\Omega}_{AB} = \tilde{d}_A \tom_B - \tilde{d}_B \tom_A$. 

The above discussion applies to general null surfaces, however stationary horizons have a geometry which is invariant in time. 
 For Killing horizons that invariance must include the spacetime in some neighbourhood of 
$\Delta$, while the slightly more general isolated horizons only require an invariant intrinsic and extrinsic horizon geometry. In either case
$\ell^a$ acts as a Killing vector on the horizon. Thus for the intrinsic metric
\bea
\Lie_\ell \tq_{AB} = 2 \tilde{k}^{(\ell)}_{AB} = 0 \, , \label{klzero}
\eea
or, decomposing this into trace and trace-free parts, the expansion $\tilde{\theta}_{(\ell)}$ and shear $\tilde{\sigma}^{(\ell)}_{AB}$ both vanish. The time 
invariance of the rest of the geometry then amounts to: 
\bea
\Lie_\ell \tilde{k}^{(n)}_{AB} = 0 \; , \; \; \Lie_\ell \tom_A = 0 \;  \mbox{and}  \; \;  \Lie_\ell \kappa_{(\ell)} = 0 \, . \label{TimeInvar}
\eea

The time invariance of $\kappa_{(\ell)}$and $\tom_A$ is sufficient to imply that $\kappa_{(\ell)}$takes a fixed value over all of $\Delta$ \cite{afk}. Identifying $\kappa_{(\ell)}$
as a surface gravity, this is the \emph{zeroth law of isolated horizon mechanics}. There is enough freedom in the structure that we have imposed to rescale
$\kappa_{(\ell)}$ by a constant\footnote{It is always possible to find some general rescaling $\ell \rightarrow f \ell$ of the null vectors for which $\kappa_{(\ell)}$ vanishes. However 
(\ref{TimeInvar}) constrains the allowed rescalings to be of the form
 $\ell \rightarrow c \ell$ for some constant $c$, so that $\kappa_{(\ell)} \rightarrow c \kappa_{(\ell)}$. 
For Killing horizons in an asymptotically flat space, this final bit freedom is eliminated by 
requiring that the evolution Killing vector field take a standard value at infinity. In the absence of such a reference, the surface gravity is only fixed up to 
a scalar multiple.}, however it is clear that in the special case $\kappa_{(\ell)}= 0$ such rescalings are 
irrelevant. In this case we say that the horizon is \emph{extremal} (a discussion of how this relates to other notions of extremality can be found in \cite{extremal}).  

\subsection{Electromagnetic properties}
From (\ref{klzero}), $\Lie_\ell \theta_{(\ell)} = 0$ and the Raychaudhuri equation, it follows that $T_{ab} \ell^a \ell^b = 0$ and so the electromagnetic field 
tensor on $\Delta$ can be written as
\bea
F_{ab} = \tilde E_\perp (\ell_a n_b - n_a \ell_b) +  \tilde B_\perp \tilde{\epsilon}_{ab} + 
(\tilde{X}_a \ell_b - \tilde{X}_b \ell_a) \, , \label{F}
\eea
for some functions $E_\perp$ and $B_\perp$, the  area form $\tilde{\epsilon}_{ab}$ on $S_v$ (induced by the metric)  and  a vector field $\tilde{X}^a \in TS_v$. As suggested by the notation, $ \tilde E_\perp$ and $ \tilde B_\perp$ 
respectively measure the electric and magnetic fluxes through $S_v$ so that the electric and magnetic charges are:
\bea
Q = \frac{1}{4 \pi} \int_{S_v} 
\tilde{\epsilon}  \tilde E_\perp  \; \; \mbox{and} \; \; 
P = \frac{1}{4 \pi} \int_{S_v} 
\tilde{\epsilon}  \tilde B_\perp \, . 
\eea
As for the geometric properties, for Killing and isolated horizons  the electromagnetic properties are invariant in time: 
$\Lie_\ell  \tilde E_\perp = \Lie_\ell  \tilde  B_\perp = 0$ and $\Lie_\ell \tilde{X}_A = 0$. In particular this means that the electric and magnetic charges are time invariant.

\subsection{Constraint equations}
It was shown in \cite{abl2} that  $\Lie_\ell k^{(n)}_{AB} = 0$ combined with the Einstein equations gives a constraint on the allowed 
horizon geometries: 
\begin{eqnarray}
\kappa_{(\ell)}& k^{(n)}_{AB} +  \tilde{d}_{A} \tom_{B} + \tom_A \tom_B 
 & =   \frac{1}{2} ( \tilde{K} 
 - \tilde Z_\perp^2 ) \tilde{q}_{ab} +  \frac{1}{2} \tOm \tep_{ab}  \label{PreEC}  
\end{eqnarray}
where 
\bea
\tilde Z_\perp^2 = \tilde E_\perp^2 + \tilde B_\perp^2 \, , 
\eea
and $\tilde{K}$ is the Gauss curvature of $S_v$ (for a two-sphere with area $4 \pi r^2$, $\tilde{K} = 1/r^2$). 

A second constraint comes from the Maxwell equations (\ref{Maxwell}) and was first derived in \cite{Lewandowski:2002ua}. It says that 
\bea
2 \kappa_{(\ell)}\tilde{X}_A = 
 (\tilde{d}_A + 2 \tom_A ) \tilde E_\perp - \tep_A^{\phantom{A} B} (\tilde{d}_B+ 2 \tom_B) \tilde B_\perp \label{PreMaxwell} \, .
 \eea

Note that for extremal horizons, $\tilde k^{(n)}_{AB}$ and $\tilde{X}_A$ are eliminated from the equations. It is from this significant simplification that the
uniqueness theorems follow.

\subsection{Extreme electrovac horizons}
\label{ExHor}

In 2002, Lewandowski and Pawlowski \cite{Lewandowski:2002ua} proved the following uniqueness theorem: \\

\textbf{Theorem I (Rotationally Symmetric Uniqueness):}  \emph{ Let $\Delta$ be an  extremal electrovac isolated horizon
with topology $\mathbb{R} \times S^2$.
 Further assume 
that both the isolated horizon geometry and electromagnetic field admit a rotational $O(2)$ symmetry generated by a vector field $\phi^a$. 
Then the constraints of the last section are satisfied if and only if the induced
metric $\tq_{AB}$, the rotation one-form $\tom_A$ and the normal electromagnetic components $\tilde E_\perp$ and $\tilde B_\perp$ coincide with 
the corresponding quantities on the event horizon of an extremal Kerr-Newman solution.}  \\



If one assumes that a non-rotating ($\tilde \Omega_{AB} = 0$) extremal horizon is axisymmetric, 
then Theorem I implies that it must be the same as an Reissner-Nordstr\"om
extremal horizon. However, following some earlier work in \cite{Chrusciel:2005pa},  it was shown in \cite{Kunduri:2008tk} that the result still holds even if we 
drop the symmetry assumption: \\

\textbf{Theorem II (non-rotating):} \emph{Let $\Delta$ be an extremal electrovac isolated horizon
with topology $\mathbb{R} \times S^2$.
Further assume that the horizon is non-rotating: $\tilde  \Omega _{AB} = 0$. Then the constraints of the last section are satisfied if and only if 
the induced metric $\tq_{AB}$, $\tom_A$, $\tilde E_\perp$ and $\tilde B_\perp$ coincide with the corresponding quantities defined on the event horizon of an 
extremal Reissner-Nordstr\"om solution. That is, horizon cross-sections are necessarily geometric spheres on which $\tom_A = 0$, and 
$\tilde E_\perp$ and $\tilde B_\perp$ are constants which satisfy
\bea
\tilde E_\perp^2 + \tilde B_\perp^2 = \frac{1}{r^2}
\eea
where $r$ is the areal radius of the sphere.  } \\

This second case is the one of interest for this paper. It should be emphasized that it makes \emph{no symmetry assumptions} about the 
geometry of the horizon cross-sections. It is the constraint equations that
force them to be spherical. Further, the derived restriction is only on the horizon geometry: the theorem allows for spherically symmetric extremal horizons
embedded in non-spherically symmetric spacetimes. 

\section{Examples of non-rotating extremal spacetimes}
\label{Examples}

In this section we construct such families of non-asymptotically flat  spacetimes which are not spherically symmetric but still contain an extremal 
Reissner-Nordstr\"om-type isolated horizon. We explicitly examine the intrinsic and extrinsic geometries of these horizons and demonstrate 
that they are consistent with  both the local and global uniqueness theorems. 

We start with Weyl-distorted Reissner-Nordstr\"om solutions which can be used to generate axisymmetric distortions of both extremal and non-extremal solutions. 
Hence we can compare and contrast behaviours between the two cases. Then we turn to conformastatic distortions. 
These can only be applied to extremal solutions, however they allow for general distortions with no symmetry assumptions. 
The families coincide for axisymmetric extremal distortions.

\subsection{Weyl-distorted Reissner-Nordstr\"{o}m}

By the Weyl ansatz, static, axially symmetric metric solutions of the Einstein-Maxwell equations can be written in the form
\be
ds^2 = - e^{2 \psi} dt^2 + e^{2(\gamma - \psi)} (d \rho^2 + dz^2) + \rho^2 e^{-2 \psi} d \phi^2 \label{weyl}
\ee
for some potentials $\psi(\rho, z)$ and $\gamma(\rho,z)$. We restrict our attention to solutions for which the electromagnetic field 
is generated by a vector potential of the form
\be
A_a = \Phi (\rho,z) [dt]_a \, . 
\ee
Then the Einstein-Maxwell equations reduce to become:
\bea
\psi_{\rho \rho} + \frac{1}{\rho} \psi_\rho + \psi_{zz}   = e^{-2 \psi} \left(\Phi_\rho^2 + \Phi_z^2 \right)  \, , \label{psi} \\
\Phi_{\rho \rho} + \frac{1}{\rho} \Phi_\rho + \Phi_{zz}   =  2 \left(  \psi_\rho \Phi_\rho + \psi_z \Phi_z   \right) \, ,  \label{Phi} \\
\gamma_\rho = \rho (\psi_\rho^2 - \psi_z^2) - \rho e^{-2\psi} (\Phi_\rho^2 - \Phi_z^2  )  \; \;  \mbox{and} \label{gammarho} \\
\gamma_z = 2 \rho \psi_\rho \psi_z - 2 \rho e^{- 2\psi} \Phi_\rho \Phi_z \label{gammaz}  \ ,
\eea
where subscripts indicate partial derivatives. 

In particular, 
both pure and distorted Reissner-Nordstr\"om black holes can be written in Weyl form \cite{gautreau,Geroch:1982bv, fk}. We now examine how
this is done. 

\subsubsection{Vacuum solutions}

Though our interest is electrovac solutions, it useful to begin with the vacuum case in order to demonstrate some important features
of these solutions in a slightly simpler setting. Hence we begin with distorted Schwarzschild solutions. 

If $\Phi = 0$ then (\ref{psi}) becomes the rotationally symmetric Laplace equation for Euclidean $\mathbb{R}^3$ in cylindrical 
coordinates. The solutions to this are well-known and can be series expanded, for example, in terms of Bessel and Neumann functions. 
Given a $\psi$ satisfying this equation, $\gamma$ is found by integrating 
 (\ref{gammarho}) and (\ref{gammaz}).

Now, the Laplace equation is linear and so superpositions of known solutions remain solutions. In particular, Schwarzschild
is generated by one such $\psi_{\!_S}$ and the family of spacetimes generated by $\psi = \psi_{\!_S} + \psi_{\!_D}$, where $\psi_{\!_D}$
is another solution of the Laplace equation, are distortions of the Schwarzschild spacetime. For completeness, 
the form of the Schwarzschild potential is given in \ref{AppA}, though we will not  need its exact form in this discussion. 

There are complications to this picture. The full Einstein equations are, of course, not linear and so full solutions cannot be so superposed. This
is reflected in  (\ref{gammarho}) and (\ref{gammaz}) which are not linear, even in the vacuum case. 
 Thus, given the Schwarzschild solution $(\psi_{\!_S}, \gamma_{\!_S})$ and a distortion $(\psi_{\!_D}, \gamma_{\!_D})$, 
the pair $(\psi_{\!_S}+\psi_{\!_D}, \gamma_{\!_S}+\gamma_{\!_D})$ generally does \emph{not} generate a solution to the Einstein equations.
$\psi_{\!_S}+\psi_{\!_D}$ is certainly a solution of the Laplace equations, but on integrating (\ref{gammarho}) and (\ref{gammaz}) the resulting 
$\gamma_{\!_{S+D}} \neq \gamma_{\!_S}+ \gamma_{\!_D}$. 
Undeterred by this fact, it will be notationally  convenient to \emph{define}
\bea
\bar{\gamma}_{\!_D} = \gamma_{\!_{S+D}} - \gamma_{\!_S} \, ,
\eea 
while always keeping in mind that $\bar{\gamma}_D \neq \gamma_D$. 

Given a pair $(\psi_{\!_S}+ \psi_{\!_D}, \gamma_{\!_S}+ \bar{\gamma}_{\!_D} )$, we transform (\ref{weyl}) into a more familiar form by applying the
coordinate transformation
\be
\rho = \sqrt{r^2 - 2 M r} \sin \theta \; \; \mbox{and} \; \; z = (r - M) \cos \theta \, . \label{Strans}
\ee
Then the distorted Schwarzschild metric takes the more familiar form
\bea
ds^2 &=& - e^{2 \psi_{\!_D}} \left( 1 - \frac{2m}{r} \right) dt^2 + e^{2(\bar{\gamma}_{\!_D} - \psi_{\!_D})} \left( \frac{dr^2}{1 - \frac{2m}{r}} + r^2 d \theta^2  \right) \\
& & + r^2 e^{-2 \psi_{\!_D}} \sin^2 \theta d \phi^2  \nonumber \, ,
\eea 
where we have rewritten $e^{2 \psi_{\!_S}}$ and $e^{2 \gamma_{\!_S}}$ in terms of $(1-2m/r)$.

Thus for any $\psi_{\!_D}$ satisfying the (axially-symmetric) Laplace equation we have a corresponding vacuum spacetime (with $\psi_{\!_D}=0$ 
corresponding to pure Schwarzschild). However, there is a second complication.  Solutions to the Laplace equation in Euclidean $\mathbb{R}^3$ 
diverge either at infinity or for $\rho = 0$. Translating back into regular coordinates, this corresponds to distortion potentials either diverging at infinity 
(in which case the spacetimes are no longer asymptotically flat) or at $r=2m$. We restrict our attention to distortion potentials that are regular at 
$r=2m$. While the corresponding spacetimes are no longer asymptotically flat, they can be 
analytically continued through $r=2m$ in more-or-less the usual way and it remains a Killing horizon\cite{Geroch:1982bv}. 
Given our focus on horizons, this is a convenient class of solutions for study\footnote{Some potentials that diverge at $r=2m$ are physically innocuous: for 
example the Schwarzschild potential itself has this property, as do distortion potentials 
that move us around through the phase space of Schwarzschild solutions. However, others will certainly induce extra singularities. 
Separating these from each other would add an extra complication to our calculations. For our purposes the distortions that are regular at $r=2m$
are sufficient and so we choose to focus on them.}.


While these solutions are not asymptotically flat, they can be used in the construction of
 \emph{non-vacuum}, asymptotically-flat distorted black hole solutions \cite{Geroch:1982bv}. With the help of spacetime 
 surgery techniques, one can retain the region of spacetime near the horizon, but replace the asymptotic region with a section 
 of an asymptotically flat spacetime along with a matter field that glues the two parts together. This mediating matter field can then be regarded
 as inducing the distortion (though straightforward in principle, the details of implementation are somewhat complicated \cite{andrew}).



%

\subsubsection{Electrovac solutions}
\label{electrovac}

With these points in mind we turn to the electrovac case and immediately note that non-asymptotically flat
spacetimes are exactly what we need: without such a violation, the global uniqueness theorems hold and so our search for non-standard
families of extremal solutions would be in vain!

However, there are now new complications. For non-vanishing electromagnetic fields, (\ref{psi}) becomes a non-linear Poisson equation. Thus, in general, superposition of solutions is no longer possible. However there is a special subset of solutions where that property 
is  recovered and we restrict our attention to solutions of that ilk. If we assume that the Coulomb potential takes the form
%
%
%
\be
\Phi = \Phi (\psi) \, , 
\ee 
then it is relatively straightforward to see that solutions of (\ref{psi}) and (\ref{Phi}) must be of the form
\bea
e^{-\psi} &=& \frac{1}{\sqrt A} \sinh (\alpha  \psi^o)  \; \; \mbox{and} \label{psiRN} \\
 \Phi &=&  \sqrt{A + e^{2 \psi}}  + C   \label{PhiRN}
\eea
where $\psi^o$ is a vacuum solution of (\ref{psi}) and $\alpha$, $A$ and $C$ are free parameters. By (\ref{PhiRN}),  (\ref{psi}) and (\ref{Phi}) 
become the same equation, while (\ref{psiRN}) transforms them  back into Laplace form\footnote{Note that the $\alpha$ doesn't 
add any extra freedom into the construction (if $\psi^o$ is a solution of the Laplace
equation then so is $\alpha \psi^o$). However it turns out to be crucial to separate this out so that we can construct a unified treatment
of non-extremal and extremal horizons. Here we follow \cite{gautreau} but depart from \cite{fk} where $\alpha$ was set to be one and so 
the extremal case was excluded.}. Superposition has been recovered, though at a different level than before: the $\psi$s appearing 
directly in the metric cannot be superposed. Instead it is the progenitor $\psi^o$s that have this property.

Finally, and almost miraculously, it turns that there is a very simple relationship between 
$\gamma$ and $\gamma^o$:
\bea
\gamma = \alpha^2 \gamma^o \, . \label{gammaRN}
\eea
Thus from any vacuum Weyl solution $(\psi^o, \gamma^o)$ and choice of the free parameters $(\alpha, A, C)$, 
we can apply (\ref{psiRN}), (\ref{PhiRN}) and (\ref{gammaRN}) to generate a corresponding electrovac solution $(\psi, \gamma, \Phi)$. 

We are interested in families of solutions that include Reissner-Nordstr\"om.
We consider solutions generated from 
\bea
e^{-\psi} &=& \frac{1}{\sqrt A} \sinh \left( \alpha  ( \psi^o_{\!_{RN}} + \psi^o_{\!_{D}} \right) ) \label{psiRN2}
\eea
with the requirement that for vanishing distortion potential $\psi^o_{\!_{D}}$ this reduces to Reissner-Nordstr\"om and in particular it 
includes the extremal limit. Requiring that this limit hold imposes restrictions on the allowed values of $\alpha$, $A$ and $C$. 

One needs to be careful in imposing this limit to avoid getting lost in an algebraic morass;
the key is to avoid ever explicitly writing out $\psi^o_{\!_{RN}}$ in cylindrical coordinates but instead work only with 
the spherical coordinate form of $e^{\psi_{RN}}$. 
Following the same pattern as (\ref{Strans}), we transform to spherical-type coordinates via
\be
\rho = \sqrt{r^2 - 2 M r + Q^2} \sin \theta \; \; \mbox{and} \; \; z = (r - M) \cos \theta \, . \label{CoordTrans}
\ee
The time is not yet right to execute the full transformation of (\ref{weyl}). For now we simply note that (\ref{CoordTrans}) leaves 
the $dt^2$ coefficient unchanged so that in spherical-type coordinates
\bea
e^{2  \psi_{\!_{RN}}} = F \equiv 1 - \frac{2M}{r} + \frac{Q^2}{r^2} \, , \label{ttcomponent}
\eea
while the Coulomb potential 
\bea
\Phi_{\!_{RN}} =  \sqrt{A + e^{2 \psi_{\!_{RN}}}}  + C = \frac{Q}{r} \, . 
\eea
Then, simple algebra tells us that\cite{gautreau}:
\bea
A = C^2 - 1 \; \; \mbox{and} \; \; C = \frac{M}{Q} \, . 
\eea

The extremal limit occurs when $C \rightarrow 1$ (from above) and to keep the $\psi$ from  (\ref{psiRN2})
well-defined in that limit, we choose $\alpha = \sqrt{A}$:
\bea 
e^{-\psi} = \frac{1}{\sqrt{A}} \sinh \left( \sqrt{A} ( \psi^o_{\!_{RN}} + \psi^o_{\!_D}  \right) \, .
\eea
Combining this with (\ref{ttcomponent}) we can eliminate all reference to $\psi_{\!_{RN}}$. To wit: 
%
\bea
e^{2 \psi} = \frac{F}{\Gamma^2} \, , 
\eea 
where 
\bea
\Gamma = \cosh \left( \sqrt{A} \psi^o_{\!_D} \right) + \sqrt{ \frac{F+A}{A} } \sinh  \left( \sqrt{A} \psi^o_{\!_D} \right) \, . 
\eea
With this result in hand, 
the time is now right to go back to the coordinate transformation (\ref{CoordTrans}) for which we find that 
\bea
e^{2 \gamma} \left( d \rho^2 + dz^2 \right)  = \mathcal{G}^2 \left( \frac{dr^2}{F} + d \theta^2 \right)
\eea
where
\bea
\mathcal{G}^2 = e^{2 \gamma}  \left(1 + \left(\frac{Q^2 A^2}{r^2 F}\right) \sin^2 \theta \right) \, . 
\eea
Then our family of distorted spacetimes may be written as
\bea
ds^2 = - \left(  \frac{F}{\Gamma^2} \right) dt^2 + \left( \frac{\Gamma^2 \mathcal{G}^2}{F} \right) dr^2 + r^2 \Gamma^2
\left(\mathcal{G}^2 d \theta^2 + \sin^2 \theta d \phi^2 \right) \, . \label{GeneralDist}
\eea

Not surprisingly, this is somewhat complicated and evaluating for specific $\mathcal{G}$ (which include $\gamma$) is 
not straightforward. In particular for $A \neq 0$ one needs an expression for $\gamma$ in order understand what is happening
at the horizon. Details for specific examples can be found in \cite{fk, frolov1, frolov2, Pilkington:2011aj}. Here we just note that for 
distortion potentials that are regular at $F(r)=0$, the spacetime may be analytically continued through that surface which also continues
to be a Killing horizon. Further, in general, distortions of the spacetime metric do indeed induce distortions of the horizon geometry: both the intrinsic and 
extrinsic geometry change and can be distorted very dramatically.

We now restrict our attention to the extremal case where $A=0$ and significant simplifications result. For this
case $\gamma = 0$  and so 
\bea
\Gamma_{ex} &=& 1 +  \sqrt{F} \psi^o_{\!_D}\; \; \mbox{and} \\
\mathcal{G}_{ex} &=& 1 \, .  \nonumber
\eea
Hence
\bea
ds_{ex}^2 = - \left(  \frac{F}{\Gamma_{ex}^2} \right) dt^2 + \Gamma_{ex}^2 \left( \frac{dr^2}{F} + r^2 \left(d \theta^2 + \sin^2 \theta d \phi^2  \right)\right) \, . \label{exWeyl}
\eea
Even a casual inspection of the metric suggests that things are very different here: at the horizon (where $F  \rightarrow 0$) the 
angular part of the metric is always a geometric sphere. However, we defer a full analysis of the geometry 
until Section \ref{HorGeom}. 


\subsection{Conformastatically distorted extremal Reissner-Nordstr\"om}

Next we consider conformastatic distortions of extremal Reissner-Nordstr\"om \cite{Gonzalez:2008rn, LoraClavijo:2010ih, conformastatic}. 
These are very close in both spirit and substance to the 
Weyl distortions that we have just considered. However while they drop our earlier axisymmetry assumption, they do not extend to the 
non-extremal case. 

%

This time the metric ansatz includes only one free function $\Psi = \Psi (\rho,z,\phi)$:
\bea
ds^2 = - e^{2 \Psi} dt^2 + e^{-2 \Psi } \left(d \rho^2 + d z^2 + \rho^2 d \phi^2 \right) \, , 
\eea
and we again assume that the electromagnetic field is generated by a static potential of the form:
\bea
A_a = \Phi(\rho,z,\phi) [dt]_a \, . 
\eea
Then the Einstein equations reduce to become:
\bea
\Psi_{\rho \rho} + \frac{1}{\rho} \Psi_\rho + \Psi_{zz} + \frac{1}{\rho^2} \Psi_{\phi \phi} = 
e^{- 2 \Psi} \left( \Phi_\rho^2 + \Phi_z^2 + \frac{1}{\rho^2} \Phi_\rho^2 \right)  \, ,  \label{preLap} \\
\Psi_i \Psi_j = e^{-2 \Psi} \Phi_i \Phi_j  \, , \label{switcheroo} 
\eea
where in (\ref{switcheroo}), $i, j \in [\rho, z, \phi]$. 

To construct the solutions we apply essentially the same strategy as for Weyl. 
For a conformastatic solution of the Einstein-Maxwell equations the electromagnetic potential necessarily takes the form\cite{conformastatic}:
\be
\Phi = \Phi (\Psi) \, . 
\ee
Then (\ref{switcheroo}) are all satisfied if 
\bea
e^{-2 \Psi} \left( \frac{d \Phi}{d \Psi} \right)^2 = 1 \; \; \Longrightarrow \Phi = \pm e^{\Psi} + C \, , \label{PhiEx}
\eea
for some constant $C$.  This leaves only (\ref{preLap}) but it is not hard to solve. Inspired by the extremal Weyl solutions, we make the
substitution
\be
e^{-\Psi} = \Psi^o \, , 
\ee
and find that  $\Psi$ solves (\ref{preLap}) if and only if $\Psi^o$ solves the general Laplace equation in Euclidean $\mathbb{R}^3$. 

We continue in the familiar vein, considering solutions generated by a Reissner-Nordstr\"om part along with a distortion: \bea
e^{-\Psi} = \Psi^o_{\!_{RN}}  + \Psi^o_{\!_D}=  \frac{1}{\sqrt{F}} + \Psi^o_{\!_D} \, .
\eea
Then given the transformation
\bea
\rho = (r-M) \sin \theta \; \; \mbox{and} \; \; z = (r-M) \cos \theta \, ,
\eea
the metric again takes the extremal Weyl form (\ref{exWeyl}):
\bea
ds^2  = - \left( \frac{F}{{\Gamma}_{ex}^2} \right) dt^2 + {\Gamma}_{ex}^2 \left( \frac{dr^2}{F} + r^2 \left(  d \theta^2 + \sin^2 \theta d \phi^2 \right) \right) \label{exWeyl2}
\eea
with
\bea
{\Gamma}_{ex} = 1 + \sqrt{F} \Psi_{\!_D}^o \, . 
\eea
though this time are no symmetry restrictions on $\Psi_{\!_D}^o$. We can set the constants in (\ref{PhiEx}) by matching against Reissner-Nordstr\"om 
for the $\Psi_{\!_D}^o = 0$ case. Then the distorted electromagnetic potential is 
\be
A_a = \left( 1- \frac{r- M}{r \Gamma_{ex} } \right)  [dt]_a \label{Expotential}
\ee

Our reason for delaying analyzing the horizon geometry is now obvious: the extremal Weyl geometries are a subset of the 
conformastatic ones and so it  is convenient to study both simultaneously.

\subsection{Horizon Geometry} \label{HorGeom}

We are ready to consider properties of the extremal horizons. Unfortunately the $(t,r,\theta,\phi)$ coordinate system is not well-defined in the only 
place we are really interested in: the horizon. We could fix this problem by switching to Eddington-Finkelstein coordinates as was done in 
\cite{fk, Pilkington:2011aj}. However this is somewhat cumbersome and so we instead choose to work 
with the surfaces of constant $t$ and $r$ along with appropriately scaled null normals to those surfaces so that in the limit $r \rightarrow r_{\mathrm{horizon}}$
the associated geometric quantities remain well-defined even if the coordinates do not. We define
\bea
\ell = \tge \frac{\partial}{\partial t} + \frac{F}{\tge} \frac{\partial}{\partial r} \; \; \; \mbox{and} \; \; \;  
n = \frac{1}{2} \left( \frac{\tge}{F} \frac{\partial}{\partial t} - \frac{1}{\tge} \frac{\partial}{\partial r} \right)  \, \label{elln}
\eea
and have already noted that the induced metric on surfaces of constant $t$ and $r$ is
\bea
dS^2 = \tge^2 r^2 \left( d \theta^2 + \sin^2 \theta d \phi^2 \right) \, . \label{SurfMet}
\eea

In Section \ref{electrovac} we noted that for distortion potentials that vanish at the $F(r)=0$, the spacetime is regular on that surface which also remains as
a Killing horizon. This result continues to hold for extremal solutions. Straightforward calculations confirm that it is an isolated horizon. For a general surface 
of constant
$t$ and $r$:
\bea
\tilde \theta_{(\ell)} = \frac{2 (r-M)^2}{r \tge^2} \left( 1 + \Psi_{\!_D}^o+ (r-M) \partial_r  \Psi_{\!_D}^o \right) \, . 
\eea
This vanishes for $r=M$ and so that surface is non-expanding in the $\ell$ direction. Further, it is tangent to $\ell$ and so clearly null. We now 
focus on this surface. There the surface gravity
\bea
\kappa_{(\ell)} = \left. - \frac{\partial}{\partial r} \left(\frac{(r-M)^2}{r^2 G}\right) \right|_{r=M} = 0 
\eea
while the inward expansion
\bea
\tilde \theta_{(n)}  & = & - \frac{1 + \Psi_{\!_D}^o(M,\theta,\phi)}{M}  \label{tnEx}
\eea
and  $ \tilde \sigma^{(\ell)}_{AB}$, $ \tilde \sigma^{(n)}_{AB}$, $\tom_A$ and $\Omega$ all vanish.  
The electromagnetic field  is specified by:
\bea
\tilde E_\perp = \frac{1}{M} \; \; \mbox{and} \; \; \tilde B_\perp = 0 
\eea
while
\bea
\tilde X_A = \frac{1}{2} \left( \partial_\theta \Psi_{\!_D}^o(M,\theta,\phi) [d\theta]_A +  \partial_\phi \Psi_{\!_D}^o(M,\theta,\phi) [d\phi]_A \right) 
\eea
and all other components vanish. Similarly the Lie derivative of any of these quantities with respect to $\ell$ (again evaluated at the horizon) 
vanishes.

Thus the surface defined by $r=M$ is null and on it $\tilde\theta_{(\ell)} = 0$ while rest of the geometry and electromagnetic field is 
invariant in time: it is an isolated horizon. Further, with $\tilde\Omega = 0$ and  $\kappa_{(\ell)} = 0$ it is both non-rotating and extremal and as such should obey Theorem II. 
It does. The induced metric $\tq_{AB}$ (and therefore the Gauss curvature $\tilde{K}$), $\tom_A$, $\tilde E_\perp$ and $\tilde B_\perp$ all match the expected Reisner-Nordstr\"om values for a horizon 
of radius $r=M$. 
The only quantities showing effects of the distortion are $\tilde\theta_{(n)}$ and $\tilde X_A$. However this is also consistent with Theorem II which does not require them to remain unaltered. 

Though we will not include the details here, we have also conducted a more comprehensive investigation of the geometric properties of 
distorted extremal horizons. Transforming to ingoing Eddington-Finkelstein-type coordinates and setting up 
an ingoing-null geodesic-adapted null tetrad\footnote{On the horizon, the Newman-Penrose $\ell$ and $n$ are equivalent to (\ref{elln})
while $m$ and $\bar{m}$ are tangent to the horizon cross-sections.} we calculated the associated Newman-Penrose quantities (see for example \cite{chandra})
and compared them with undistorted Reissner-Nordstr\"om. In accord with results given above, the quantities associated with intrinsic geometry are unchanged 
while the extrinsic (Newman-Penrose)
quantities $\mu$ (equivalent to $\theta_{(n)}$ above), $\psi_3$, $\phi_2$, $\Phi_{12}$ and $\Phi_{23}$  are all distorted.

\section{Discussion}
\label{discuss}

In this paper we have explicitly demonstrated the existence of non-Reissner-Nordstr\"om spacetimes that nevertheless contain 
extremal Reissner-Nordstr\"om isolated horizons. They evade the global uniqueness theorems by not being asymptotically flat. 
Nevertheless, the solutions are still quite interesting: the overall spacetime can be dramatically distorted while the extremal horizon
remains spherically symmetric and intrinsically unperturbed.

By contrast, distortions of the surrounding spacetime do induce corresponding distortions of non-extremal horizons. 
From (\ref{GeneralDist}) the induced metric on a general Weyl-distorted Reissner-Nordstr\"om horizon is
\be
dS^2 =  r^2 \Gamma^2 \left(\mathcal{G}^2 d \theta^2 + \sin^2 \theta d \phi^2 \right)
\ee
which is manifestly non-spherical for non-trivial $\Gamma$ and $\mathcal{G}$. Further discussion can be found in \cite{fk} and 
explicit calculations demonstrating these distortions can be found in \cite{frolov1,Pilkington:2011aj} (which focus on Schwarzschild) or 
\cite{frolov2} (which studies the interiors of charged Weyl-distorted black holes). In particular \cite{Pilkington:2011aj} explicitly constructs extreme 
quadrupole and dipole-octopole distortions of Schwarzschild black holes and shows that distortions can stretch a black hole horizon  into a long thin line or 
flatten it into a pancake. 
 
How is it that a spherically symmetric extremal horizon can coexist with an non-spherically symmetric surrounding spacetime? At one level this is 
obvious: there is a uniqueness theorem for extremal horizons but no such theorem for non-extremal ones. This is true. However, we (meaning 
at least the authors) do not have a good intuition for why this theorem should exist. 
%
%
%
It has been argued (see for 
example \cite{Amsel:2009et}) that the physical root of the uniqueness theorems is that extremal horizons are an infinite distance from 
the rest of spacetime. Thus, one would not be expect them to be influenced by distortions that are always infinitely far away. While at first thought this
may be persuasive,  it is not completely satisfactory: for example from (\ref{tnEx}) it is clear that $\tilde \theta_{(n)}$, the expansion/contraction of 
ingoing null rays crossing the horizon,  \emph{is} affected by the distortions, even for extremal black holes. Unfortunately, we currently do not have any extra 
physical insight to add to this problem. 

This paper has focussed on spherically symmetric horizons but we also mentioned the corresponding uniqueness theorems for (non-trivial) Kerr-Newman 
horizons. Given our experience here, one would assume that these can also co-exist with surroundings 
that do not have the standard Kerr-Newman geometries. Indeed the existence of non-standard (and non-stationary) spacetimes containing extremal Kerr 
horizons was demonstrated in \cite{dainKerr}. However, as far as we know an analogous construction to that in the current paper has not yet been 
performed for stationary distorted Kerr-Newman spacetimes. It would be particularly interesting to investigate whether or not one could construct a 
non-axially symmetric spacetime which still contained a standard (axially symmetric) extremal Kerr-Newman horizon. However, this will be left to future work. 

%
%


\ack This work was financially supported by the Natural Sciences and Engineering Research Council of Canada. \\

\appendix

\section{Weyl potentials for Reissner-Nordstr\"om black holes} \label{AppA}

In order to demonstrate why we studiously avoided working with the generating potentials for Schwarzschild and Reissner-Nordstr\"om in the
main text, we list them in this appendix (for $\alpha=1$ -- the extension to other choices for $\alpha$ is trivial). 

The Weyl metric functions $\{\psi,\gamma\}$ are:
\bea \label{RN metric functions}
\psi_{\!_{RN}} & =& \frac{1}{2}\ln \left( \frac{L^2-(M^2-Q^2)}{(L+M)^2} \right) \\
 \gamma_{\!_{RN}}& =& \frac{1}{2}\ln \left( \frac{L^2-(M^2-Q^2)}{l_+  l_-} \right) \nonumber
\eea
where
\bea
L &=& (l_+ + l_-)/2 \; \; \mbox{and} \\
 l_\pm & =& \sqrt{\rho^2+(z\pm \sqrt{M^2-Q^2})^2} \nonumber
\eea
For these functions the Weyl metric (\ref{weyl}) becomes
\bea
\label{RN in Weyl canonical}
ds^2 & =&  -\frac{L^2-(M^2-Q^2)}{(L+M)^2}dt^2+\frac{(L+M)^2}{l_+  l_-}(d\rho^2+dz^2) \\
& & +\frac{(L+M)^2}{L^2-(M^2-Q^2)}\rho^2 d\phi^2 \; , \nonumber
\eea
which reduces to the standard form via the coordinate transformations:
\be
 \rho = \sqrt{ r^2 - 2Mr + Q^2} \sin \theta \; \; \mbox{and} \; \;  z = (r-M) \cos \theta  
\ee
and $L = r - M$. 

The Weyl functions and transformations go smoothly to the extremal limit, in which things become much simpler. Then
\bea
\psi_{\!_{ERN}}  = \ln \left( \frac{L}{L+M} \right) \; \; \mbox{and} \; \; \gamma_{\!_{ERN}}=0 \, . 
\eea
where
\bea
L = l_+ = l_- = \sqrt{\rho^2 + z^2} \, ,  
\eea
so that the Weyl metric becomes:
\begin{equation}\label{ERN in Weyl canonical}
ds^2=-\frac{L^2}{(L+M)^2}dt^2+\frac{(L+M)^2}{L^2}\,\big(d\rho^2+dz^2+\rho^2d\phi^2\big)\:.
\end{equation}
The coordinate transformations which reduce this to standard form also simplify:
\bea
z = (r-M) \cos \theta \; \; \mbox{and} \; \; \rho = (r-M)\sin \theta \, ,
\eea
and $L = r-M$. \\

\end{document}